\documentclass[twocolumn,showpacs,preprintnumbers,amsmath,amssymb]{revtex4}

\usepackage{graphicx}
\usepackage{dcolumn}
\usepackage{bm}

\begin{document}

\preprint{\it Phys. Rev. Lett. {\bf 97},  026601 (2006)}

\title{Quantum transport of slow charge carriers
in quasicrystals and correlated systems}

\author{Guy Trambly de Laissardi\`ere}
\affiliation{Laboratoire de Physique Th\'eorique et Mod\'elisation,
CNRS/Universit\'e de Cergy-Pontoise,\\
2 av. A. Chauvin, 95302 Cergy-Pontoise, France}%

\author{Jean-Pierre Julien}
\author{Didier Mayou}
\affiliation{
Laboratoire d'Etudes des Propri\'et\'es Electroniques
des Solides, CNRS,\\
BP 166, 38042 Grenoble Cedex 9, France
}%

\date{\today}

\begin{abstract}
We show that  the semi-classical model of conduction 
breaks down if the mean free path of charge carriers  is smaller 
than a typical extension of their wavefunction. 
This situation is realized for   sufficiently slow charge carriers 
and leads to  a transition from a metallic like to an  insulating like 
regime when scattering by defects  increases. 
This explains the unconventional conduction properties 
of quasicrystals and related alloys. 
The conduction properties of some heavy fermions or polaronic systems, 
where charge carriers are also slow, 
present a deep analogy.
\end{abstract}

\maketitle


The semi-classical Bloch-Boltzmann theory, which plays a fundamental role 
in our understanding of conduction in solids, has  limitations such as for 
example magnetic and electric breakdown \cite{AshcroftMermin} or quantum 
interferences in the diffusive regime \cite{Lee85} that are 
well known and have been intensively studied. 
The present work focuses on another limitation that has received 
little attention and that occurs when  charge carriers have 
sufficiently small velocities.

Indeed the semi-classical theory of conduction in crystals  
is based on the concept  of a charge carrier wave-packet 
propagating at a velocity  $V = (1/\hbar) \partial E_n(k)/ \partial k$,
where ``$E_n(k)$'' is the dispersion relation for band $n$ and wavevector k. 
The validity of the wave-packet concept requires that  the extension 
$L_{\mathrm{wp}}$ of the wave-packet  of the charge carrier is smaller 
than the distance $V \tau $ of traveling between two scattering events 
separated by a time $\tau $. On the contrary, if   
$V \tau < L_{\mathrm{wp}}$, 
a condition that can be realized by sufficiently slow charge carriers,  
the semi-classical model breaks down. 
Here we report on a quantum
theory that allows  to treat on the same footing the  standard
 regime  where the semi-classical approach is valid and the regime of 
slow charge carriers.
We find  that when the scattering time $\tau $  
decreases a transition can occur between a metallic regime 
for $V \tau > L_{\mathrm{wp}}$
and an insulating like regime for $V \tau < L_{\mathrm{wp}}$.
As an example  we consider specifically a complex metallic alloy:
the $\alpha$-AlMnSi phase. 
Ab-initio band structure calculations
show that the samples of this system, 
that have been studied experimentally, 
are in the small velocity regime 
$V \tau < L_{\mathrm{wp}}$.
This explains their unconventional conduction properties.
The $\alpha$-AlMnSi phase is structurally related to the
icosahedral quasicrystalline
phase AlMnSi and shares many similar conduction properties
with other icosahedral phases such as  AlCuFe and AlPdMn and their
crystalline approximants
\cite{Klein91,Poon92,Berger94}. 
Thus the present work  is relevant for these systems too and gives a strong
insight in the so far unexplained properties of this class of materials. 
The concepts developped here open also a new  insight in the physics 
of correlated systems. Indeed recent studies of some heavy fermions 
or polaronic systems 
\cite{Vidhyadhiraja05, Fratini03, Fratini05},  
where charge carriers are also slow, show that their conduction properties 
present a deep analogy with those described here. 
In particular a transition is observed from a metallic like regime 
at low temperature (weak scattering) to an insulating 
like regime at higher temperature (stronger scattering).


We come now to  the treatment of the  conductivity
of  independent electrons.
According to the Einstein relation the conductivity
$\sigma$ depends on the diffusivity $D(E)$ of electrons 
of energy $E$
and the density of states $n(E)$ (summing the 
spin up and spin down contribution).
We assume that $n(E)$
and $D(E)$ vary weakly on the thermal energy scale $kT$,
which is justified here.
In that case, the Einstein formula
writes
$\sigma = e^2 n(E_{\mathrm{F}})D(E_{\mathrm{F}})$,
where $E_{\mathrm{F}}$ is the chemical potential
and $e$ is the electronic charge.
The temperature dependence of $\sigma$ is 
due to the variation of the diffusivity $D(E_F)$ with
temperature.
The central quantity is thus the diffusivity which 
is related to quantum diffusion.
 
For a time independent hamiltonian, 
the quantum diffusion of states at energy $E$
can be measured through the square
of their spreading defined as :
\begin{equation}
\Delta X^2(E,t) = \Big\langle (X(t) - X(0))^2 \Big\rangle_E
\end{equation}
$X(t)$ is the Heisenberg representation of the position of
the electron along a chosen axis ``$x$''. 
$\big\langle ~\big\rangle_E$ means an average over
all states with energy equal to $E$.
The diffusivity
is given by
\begin{equation}
D(E) = \frac{1}{2} \, {\rm lim}_{t \rightarrow \infty}
\frac{{\rm d} \Delta X^2(E,t)}{{\rm d}t}.
\label{Eq_D_lim_DX}
\end{equation}

$\Delta X^2(E,t)$
can be computed trough the velocity correlation
$C(E,t)=\Big\langle V_x(t)V_x(0) + V_x(0)V_x(t) \Big\rangle_E$.
$C(E,t)$ measures the correlation between the velocities at
time $t_0$ and $t_0+t$ and is independent of $t_0$. It can be shown that
\cite{Mayou00} :
\begin{equation}
\frac{{\rm d} \Delta X^2(E,t)}{{\rm d}t} = \int^{t}_{0} C(E,t') {\rm d} t',
\label{eq_X2_C}
\end{equation}
which allows to compute  $\Delta X^2(E,t)$ in terms of
the velocity correlation function $C(E,t)$ with the initial
condition $\Delta X^2(E,t=0)=0$. For a perfect crystal $C(E,t)$ is given by
$C(E,t)=\big\langle C(n \vec k,t) \big\rangle_E$
where  
$|n \vec k \rangle$ are Bloch states with energy 
$E_n(\vec k) = E$ 
and:
\begin{equation}
C(n \vec k,t) =
2 \sum_m |V_{n,m}(\vec k)|^2 {\rm cos}\Big((E_n(\vec k)-E_m(\vec k))
\frac{t}{\hbar}\Big).
\label{eq_Cnk_Vx2}
\end{equation}
$V_{n,m}(\vec k)$ are matrix elements of the 
velocity operator $V_x$ between  $|n \vec k \rangle$  
and $|m \vec k \rangle$.
In the sum (\ref{eq_Cnk_Vx2}), 
the diagonal elements ($m=n$) are time independent, whereas
the other terms depend on $t$.
In the perfect crystal $\Delta X^2(E,t)$,
which is linearly related to $C(E,t)$ by (\ref{eq_X2_C}),
is the sum of two terms:
\begin{equation}
\Delta X^2(E,t) = V^2 t^2 + \Delta X_{\mathrm{NB}}^2(E,t).
\label{Eq_DeltaX2}
\end{equation}
The first term is the ballistic contribution where ``$V^2$''
is the average of $|V_{n,n}(\vec k)|^2$ over all states
$|n \vec k \rangle$
at the energy $E$. 
The semi-classical theory
is equivalent to take into account only this first term.
The second term in (\ref{Eq_DeltaX2}),
$\Delta X^2_{\mathrm{NB}}(E,t)$,  is
a non ballistic (non Boltzmann) contribution.
It is
due to the non diagonal
elements
of the velocity operator
($m\neq n$, in (\ref{eq_Cnk_Vx2}))
and it describes
the spreading of the wavefunction
in the unit cell. 
It can be shown that $\Delta X^2_{\mathrm{NB}}(E,t)< L^2$, 
where  $L$ is
the unit cell length along the $x$ axis.
In presence of a disorder characterized by 
a scattering time $\tau$ 
such that the non ballistic term dominates i.e.  
$\Delta X^2_{\mathrm{NB}}(E,\tau) > V^2 \tau^2$, 
the semi-classical theory breaks down and
the conductor
is a non standard metal.

\begin{figure}[]

\includegraphics[width=8cm]{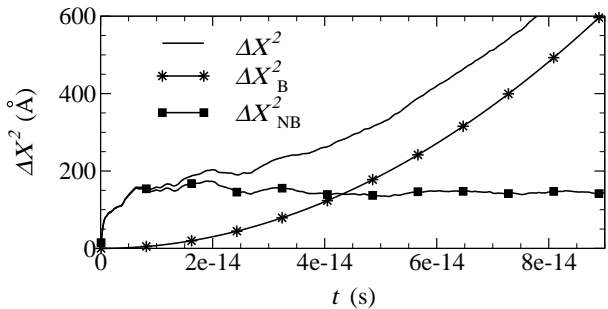}

\caption{\label{fig_X2}
Square spreading
$\Delta X^2$
of electrons states with Fermi energy
$E_{\mathrm{F}}$ versus time $t$,
in the cubic approximant 
$\alpha$-Al$_{69.6}$Si$_{13.0}$Mn$_{17.4}$. 
$\Delta X^2$ is the sum of a Boltzmann term,
$\Delta X^2_{\mathrm{B}} = V_{\mathrm{F}}^2 t^2$,
and a non Boltzmann term
$\Delta X^2_{\mathrm{NB}}$.}
\end{figure}
An example of such a crossover from a standard (ballistic) behavior to
a non standard (non ballistic) behavior is shown on  Fig. \ref{fig_X2} 
in the case of the Fermi energy states 
of the complex phase $\alpha$-AlMnSi
(see below for more details).
The non Boltzmann term
$\Delta X^2_{\mathrm{NB}}(t)$
increases very rapidly and saturates
to a maximum value of the order of  the square size of the unit cell.
At small time $t$,
the standard Boltzmann contribution,
$\Delta X^2_{\mathrm{B}}(t) = V^2 t^2$,
is smaller due to a very small velocity
$V$
of the electrons.
Thus $\alpha$-AlMnSi is a non conventional metal at these time scale
i.e. when the scattering time $\tau$ is less than $\sim 3-4\, 10^{-14}$\,s.
By contrast in pure Al, $\Delta X^2_{\mathrm{B}}$ is typically
$10 ^3$ times larger and $\Delta X^2_{\mathrm{NB}}$ is about 10 times
smaller. Therefore the non Boltzmann  contribution
is negligible at these time scales and Al is a standard metal. 

For a system with disorder (static disorder and/or temperature dependent scattering),
we compute the diffusivity 
within a relaxation time approximation
\cite{Mayou00} :
\begin{equation}
C'(E,t) = C(E,t) ~{\rm exp}(-t/\tau),
\label{Eq_Approx_Relax_t}
\end{equation}
where $C$ is the velocity correlation function of the perfect crystal
and $C'$ is that of the system with defects. 
A relaxation time $\tau$ 
is defined, which meaning is the following. 
After (\ref{Eq_Approx_Relax_t}) and (\ref{eq_X2_C}), at a time scale 
$t < \tau$ the propagation of wavefunctions is not affected by 
disorder and  is that of the perfect crystal, but  
at a time scale  $t > \tau$ the propagation becomes  diffusive 
due to the disorder.
(\ref{eq_X2_C}), (\ref{Eq_Approx_Relax_t}) 
and (\ref{Eq_D_lim_DX}) 
lead to the expressions of
the diffusivity $D$ and  of the dc-conductivity $\sigma$ of 
the system with defects:
\begin{eqnarray}
D &=& V^2 \tau + \frac{L^2(\tau)}{\tau}, \label{Eq_diffusivite}\\
\sigma  &=& e^2 n(E_{\mathrm{F}})V^2 \tau
+ e^2 n(E_{\mathrm{F}})\frac{L^2(\tau)}{\tau}.
\label{Eq_sigma}
\end{eqnarray}
Thus $D$ and  $\sigma$ have two parts.
The first ones, named in the following
$D_{\mathrm{B}}$ and $\sigma_{\mathrm{B}}$,
are the usual Bloch-Boltzmann results, and the
second ones,
named $D_{\mathrm{NB}}$ and $\sigma_{\mathrm{NB}}$,
are due to the non Boltzmann contributions.
$L^2(\tau)$ is a proper average of 
$\Delta X^2_{\mathrm{NB}}(E_{\mathrm{F}},t)$
on a time scale $\tau$. 
The second term  in the right hand side of (\ref{Eq_sigma})
is equivalent to 
the conductivity of Anderson insulators 
in the Thouless regime
\cite{Thouless77,Ovadyahu86}.
Indeed in this regime the wavefunction
spreads between two inelastic scattering events but the spreading
saturates to the localization length.
Here the localization length is replaced by the
large time limit of
$L(t)$ which, as can be shown, is bounded by the unit cell
length $L$.


We come now to a detailed analysis of the unusual transport properties
of the cubic 1/1 icosahedral approximant AlMnSi
\cite{Sugiyama98}. These results will be compared with those of
pure Al (f.c.c.),
cubic Al$_{12}$Mn \cite{PMS05}
and orthorhombic Al$_6$Mn \cite{PMS05} crystals,
which have a metallic behavior.
For all those systems, the electronic structure 
calculation have been performed
\cite{Fujiwara89,Zijlstr03,PMS05}
by using the self-consistent Tight-Binding
Linear Muffin Tin orbital (TB-LMTO)
method
\cite{Andersen75}.
Starting from the computed eigenstates $|n \vec k \rangle$,
we use equations (\ref{eq_X2_C})
and (\ref{eq_Cnk_Vx2}) to compute
the velocity correlation function without disorder,
$C(E,t)$ and the square of the spreading  $\Delta X^2(E,t)$.
The average
$\big\langle ~\big\rangle_{E}$ is obtained
by taking the eigenstates of each $\vec k$ vector with an energy
$E_n(\vec k)$ such as:
$E- \Delta E /2 < E_n(\vec k) < E+
\Delta E/2 $.
$\Delta E$ is an energy resolution in the calculation.
The number of k-points per first Brillouin zone
included in this average calculation
is $N_k$.
For a given $\Delta E$, $N_k$ is increased until
$C$ does not depend significantly on $N_k$.
Therefore, there is not adjustable parameter in our calculation.
For Al (f.c.c.), Al$_{12}$Mn, Al$_6$Mn
and
$\alpha$-Al$_{69.6}$Si$_{13.0}$Mn$_{17.4}$,
$\Delta E$ is equal to
0.272; 0.272; 0.272; 0.0272~eV,
respectively,
and $N_k$ is equal to 1,728,000; 64,000, 18,000; 32,768; respectively.


\begin{figure}[]
\includegraphics[width=7cm]{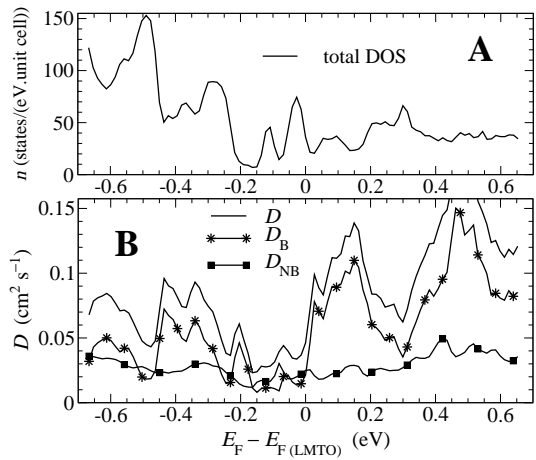}
\caption{\label{fig_DOS_sig_dif}
({\bf A}) LMTO total DOS $n(E_{\mathrm{F}})$
and ({\bf B}) diffusivity $D(E_{\mathrm{F}})$ in the
cubic approximant $\alpha$-Al$_{69.6}$Si$_{13.0}$Mn$_{17.4}$
($\tau = 3\,10^{-14}$\,s).
$D$ is the sum of a Boltzmann term, $D_{\rm B}$, and
a non Boltzmann term, $D_{\rm NB}$ 
(see equation (\ref{Eq_diffusivite})).}
\end{figure}

For the  $\alpha$-AlMnSi phase, 
we use the experimental atomic structure
\cite{Sugiyama98}
and the Si positions proposed by E. S. Zijlstra and S. K. Bose
\cite{Zijlstr03}
to calculate transport properties in an $\alpha$-phase with composition:
$\alpha$-Al$_{69.6}$Si$_{13.0}$Mn$_{17.4}$.
This phase contains 138 atoms in a cubic unit cell:
96 Al atoms, 18 Si atoms, and 24 Mn atoms.
In Fig. \ref{fig_DOS_sig_dif}, the total DOS $n$ of the 
$\alpha$-AlMnSi phase
is presented versus the energy. Its total density of
states is characterized \cite{Fujiwara89}
by a depletion near the  Fermi energy
$E_{\mathrm{F}}$, called pseudo-gap,  
which is observed experimentally.
This 
is due
to a Hume-Rothery mechanism \cite{Fujiwara91,PMS05} 
of band energy minimization 
that is strong in quasicrystals 
and their approximants \cite{Fujiwara91}.
Following the Hume-Rothery condition, 
it is expected that the most realistic value of
$E_{\mathrm{F}}$ in the actual $\alpha$-phase corresponds to
the minimum of the pseudo-gap, i.e.
$E_{\mathrm{F}}-E_{\mathrm{F(LMTO)}} = -0.163$\,eV for
our calculation.
Within a relaxation time approximation the diffusivity
$D(E,\tau)$ is calculated.
The  $D_{\mathrm{B}}$ values
(Fig. \ref{fig_DOS_sig_dif})
are similar in magnitude to those
obtained by Fujiwara {\it et al.}
\cite{Fujiwara89}
for the idealized approximant $\alpha$-Al$_{114}$Mn$_{24}$ approximant.
$D_{\mathrm{NB}}$ is almost independent
on $E$, whereas the
$D_{\mathrm{B}}$ values depend strongly on $E$
and is particularly small in the pseudo-gap.


\begin{figure}[]
\includegraphics[width=8cm]{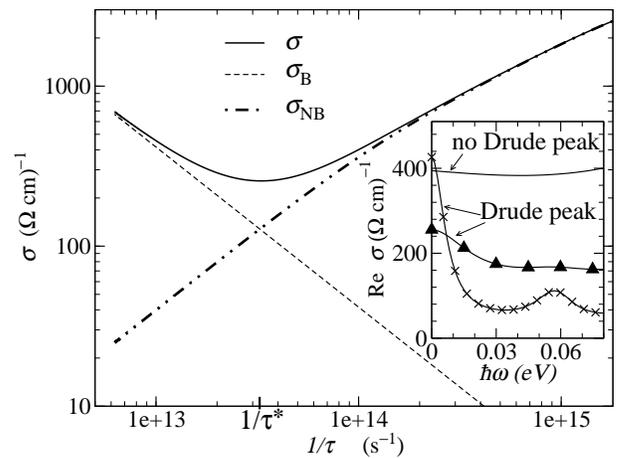}
\caption{\label{fig_sigma_omega}
Ab-initio dc conductivity, $\sigma$,
in cubic approximant $\alpha$-Al$_{69.6}$Si$_{13.0}$Mn$_{17.4}$.
$\sigma$ is the sum of a Boltzmann term, $\sigma_{\rm B}$, and
a non Boltzmann term, $\sigma_{\rm NB}$
(see equation (\ref{Eq_sigma})).
Inset: Real part, $\rm{Re}\,\sigma(\omega)$, 
of the 
optical conductivity
for
three $\tau$ values.
$\omega$ is the pulsation.
Simple line, $\tau = \tau^*/3$;
line with triangles, $\tau=\tau^*= 3.07 \, 10^{-14}$\,s;
line with crosses, $\tau = 3 \tau^*$.}
\end{figure}

We present now the results of the conductivity calculations of the 
$\alpha$-AlMnSi phase
assuming the value of the Fermi energy given above i.e. 
$E_{\mathrm{F}}-E_{\mathrm{F(LMTO)}} = -0.163$\,eV.
The Fig. \ref{fig_sigma_omega} shows the predicted static conductivity
(dc conductivity)
versus the inverse scattering time.
In the case of  $\alpha$-Al$_{69.6}$Si$_{13.0}$Mn$_{17.4}$
two regimes appear clearly: 
the metallic regime (Boltzmann regime) at large scattering time,
$\tau >\tau^*$, 
and the insulating like regime (non Boltzmann regime)
at small scattering time, $\tau <\tau^*$. 
$\tau^*= 3.07 \, 10^{-14}$\,s is defined as the time for which 
the Boltzmann and non-Boltzmann contributions are equal.
As expected from our model, 
$\sigma_{\mathrm{NB}}$ is almost proportional to $1/\tau$. 
Therefore, in the non Boltzmann regime, the conductivity 
increases with disorder 
as observed experimentally.
For realistic $\tau$ values,
$\tau <  \tau^*$
\cite{Mayou93}, $\sigma_{\mathrm{NB}}$ dominates and
$\sigma$ increases when $1/\tau$ increases i.e. when
defects or temperature increases.
$\sigma$ varies from 250~($\Omega$\,cm)$^{-1}$ for
$\tau = 3.3\, 10^{-14}$\,s,
to 2000~($\Omega$\,cm)$^{-1}$ for
$\tau = 10^{-15}$\,s.
This is consistent with experimental results
in $\alpha$-AlMnSi:
$\sigma(4~K) \simeq 200$~($\Omega$\,cm)$^{-1}$
and $\sigma(900~K) \simeq 2000$~($\Omega$\,cm)$^{-1}$ 
and with standard estimates for the scattering time 
in these systems \cite{Berger94}.
Furthermore for $\tau$ equals a few~$10^{-14}$\,s,
i.e. when the Boltzmann term is negligible, the mean free path is given
by the square root of the saturation value of
$\Delta X^2_{\mathrm{NB}}$ 
and is of the order of 15\,$\rm \AA$. This is in agreement with estimates
in the literature
\cite{Berger94}. 
As discussed in
\cite{Mayou00}
this means also that the systems is far from the
Anderson transition despite its low conductivity.
Within the relaxation time approximation used here, 
the optical conductivity is
also the sum of two terms.
One is the Boltzmann contribution
(diagonal elements of
the velocity operator) which gives rise to the so-called Drude peak
and the other is the non Boltzmann conductivity (off diagonal elements 
of the velocity operator).
As shown in the inset of Fig \ref{fig_sigma_omega} 
a Drude peak can be identified in the Boltzmann  regime,
$\tau >\tau^*$, whereas in the non Boltzmann regime,
$\tau < \tau^*$, the Drude peak disappears.
This absence of the Drude peak is experimentally demonstrated for
several icosahedral phases
\cite{Homes91_Burkov94}.
We note also that the physics of the non Boltzmann regime is similar to that
obtained with anomalous diffusion in the subdiffusive case
\cite{Mayou00,Fujiwara96,Roche96,Roche98,Bellissard03}.

\begin{figure}[]
\includegraphics[width=7cm]{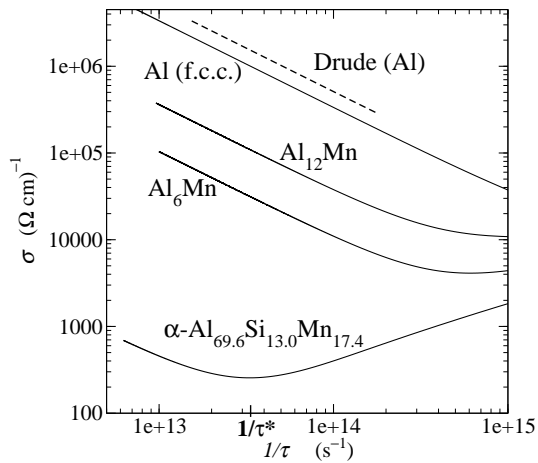}
\caption{\label{Fig_sig}
Ab-initio electrical conductivity $\sigma(E_{\mathrm{F}})$
versus inverse scattering time
$1/\tau$ (logarithmic scales):
in  Al,
Al$_{12}$Mn,
Al$_6$Mn,
and in cubic approximant $\alpha$-Al$_{69.6}$Si$_{13.0}$Mn$_{17.4}$.
In Al,
the Boltzmann term dominates, and the model is compatible
with a simple Drude model (dashed line)
\cite{AshcroftMermin}.
}
\end{figure}

Finally Fig. \ref{Fig_sig} exhibits the fundamental difference between 
the  $\alpha$-AlMnSi phase and a standard metal like Al (f.c.c). 
Al$_{12}$Mn and Al$_6$Mn phases are somewhat intermediate and show 
the first signs of an insulating like regime at the strongest 
values of the scattering rate.


To conclude,
a regime of charge carriers with small velocities has been identified,
where quantum effects play a fundamental role and lead to a transition 
from a metallic like behavior
to an insulating like behavior when the scattering rate increases.
Quasicrystals and related complex metallic alloys give an example
of this insulating like regime.
Other systems are known where the charge carriers velocity
is small, either in all directions or in specific directions, 
thus we expect that the insulating like regime could be observed 
in other systems. For interacting electrons, a similar regime
should also  exist provided that  the mean free path is smaller
than a characteristic extension  of the quasiparticle wavefunction,
which means that the Fermi liquid picture breaks down.
Several interacting electrons systems are good candidates
for this small velocity regime. Recent results
suggest this picture for a system like CeB$_6$ which is a heavy fermion
system with large mass and thus small velocity.
At low temperature $T$ the Fermi
liquid description is valid but at higher temperature 
($T>T^*$)
an incoherent insulating like regime is reached
in which the resistivity decreases with
increasing temperature.
Within a Dynamical Mean Field Theory (DMFT) calculation
\cite{Georges96}
a disappearance of the
Drude peak in the incoherent regime i.e. above $T^*$ is predicted
\cite{Vidhyadhiraja05}.
Recent theoretical results for the small polaron problem, 
with narrow polaronic
band and thus small velocity, have been obtained also with the
DMFT.
They exhibit an analogous behavior
\cite{Fratini03}.
The static resistivity (dc resistivity)
increases with increasing temperature  up to some
$T_0$ and then decreases.
Again in the incoherent regime
above $T_0$ the Drude peak disappears
\cite{Fratini05}.
A theoretical analysis similar to the discussion given
here can be made also in the DMFT.
It should give an additional insight in the quantum transport of
correlated charge carriers with small velocities.

The computations have been performed at the
Service Informatique Recherche (S.I.R.),
Universit\'e de Cergy-Pontoise \cite{Condor}.
GTL thanks Y. Costes and J.C. Baccon, S.I.R.,
for computing assistance.

\end{document}